\documentclass{aa}

\usepackage{graphicx,times}

\begin{document}

\title{XMM-Newton observations of the cluster of galaxies S\'ersic 159-03}

\author{ J.S. Kaastra \inst{1}
         \and
         C. Ferrigno \inst{1}
         \and
         T. Tamura \inst{1}
         \and
         F.B.S. Paerels \inst{2}
         \and
         J.R. Peterson \inst{2}
         \and
         J.P.D. Mittaz \inst{3}
         }
  
\offprints{J.S. Kaastra}
\mail{J.Kaastra@sron.nl}

\institute{ SRON Laboratory for Space Research
              Sorbonnelaan 2, 3584 CA Utrecht, The Nether\-lands 
              \and
              Astrophysics Laboratory, Columbia University, 
              550 West 120th Street, New York, NY 10027, USA
              \and
              Mullard Space Science Laboratory, University College,
              London, Holmbury St. Mary, Dorking, Surrey, RH5 6NT, UK
              }

\date{Received  / Accepted  }

\abstract{The cluster of galaxies S\'ersic~159$-$03\ was observed with the
XMM-Newton X-ray observatory as part of the Guaranteed Time program.  X-ray
spectra taken with the EPIC and RGS instruments show no evidence for the strong
cooling flow derived from previous X-ray observations.  There is a significant
lack of cool gas below 1.5~keV as compared to standard isobaric cooling flow
models.  While the oxygen is distributed more or less uniformly over the
cluster, iron shows a strong concentration in the center of the cluster,
slightly offset from the brightness center but within the central cD galaxy.
This points to enhanced type Ia supernova activity in the center of the cluster.
There is also an elongated iron-rich structure extending to the east of the
cluster, showing the inhomogeneity of the iron distribution.  Finally, the
temperature drops rapidly beyond 4\arcmin\ from the cluster center.
}

\maketitle

\section{Introduction}

The visible mass in clusters of galaxies is dominated by hot diffuse gas.  Due
to the high temperature of the gas it is predominantly visible in the X-ray
band.  X-ray spectroscopy is the key tool to understand the physics of this gas
and its role in the structure and evolution of the cluster.  Since the gas is
extended, spatially resolved spectroscopy is required.  Until recently
medium-resolution spectroscopy using Gas Scintillation Proportional Counter
(GSPC) or Charge-Coupled Device (CCD) technology was limited to a spatial
resolution of a few arcminutes (ASCA and BeppoSAX).  Therefore only in a
handfull of nearby clusters the core could be resolved spatially.  While other
instruments aboard the Einstein and Rosat satellites had better spatial
resolving power, they either lacked bandwidth or spectral resolving power.  High
spectral resolution data of clusters have only been obtained for the bright core
of the Virgo cluster with the Einstein FPCS detector (Canizares et al.
\cite{canizares}).  With XMM-Newton it is now possible to combine
high-resolution spectroscopy of moderately extended sources using the Reflection
Grating Spectrometer (RGS) with high-sensitivity, medium (CCD-type) spectral
resolution imaging with the European Photon Imaging Camera (EPIC) on spatial
scales down to a few arcseconds.

Here we report the XMM-Newton observation of the rich cluster of galaxies
\object{S\'ersic~159$-$03}, discovered by S\'ersic (\cite{sersic}), also named
Abell S~1101.  The cluster shows within the central 2\arcmin\ a cooling flow of
230~M$_{\sun}$/year (Allen \& Fabian \cite{allen}), centered on the dominant cD
galaxy \object{ESO 291$-$9}.  The only redshift measurement of the cluster
($z$=0.0564) is from this central galaxy (Maia et al.  \cite{maia}).  Using
$H_0$=50~km\,s$^{-1}$\,Mpc$^{-1}$ and $q_0=0.5$ we then have a luminosity
distance of 343~Mpc and an angular size distance of 307~Mpc.  We adopted a
galactic column density of $1.79\times 10^{24}$~m$^{-2}$ (Dickey \& Lockman
\cite{dickey}, using NASA's w3nH tool).

\section{Observations}

The observations were obtained on May 11, 2000.  Data processing was done using
the development version of the Science Analysis System (SAS) of XMM-Newton.  Due
to enhanced and variable background, in particular at the end of the observation
we only used 32000~s for EPIC and 36000~s for RGS.  Background subtraction for
both EPIC and RGS was done using an exposure of the Lockman hole, with similar
data selections and the same extraction regions as for the cluster, scaling
according to the exposure time.

In the EPIC data we excluded the 7 strongest point sources both in the
S\'ersic~159$-$03\ and Lockman hole field.  The thin filter was used for both
MOS cameras.  In both RGS instruments, we ignored data below $\sim 10$\AA\ due
to problematic background subtraction.

\section{EPIC analysis}

We first fitted the spectrum using isothermal models. We determined the
abundances of the elements with respect to iron in the 1--6\arcmin\
range, where the spectrum has good statistics and the effects of a
temperature drop in the center and outer regions appear to be small
(see later). We determine abundances relative to solar, where we take
the solar abundances of Anders \& Grevesse (\cite{anders}), and then
express all abundances relative to the iron abundance, the best
determined abundance with a relative error of only 7.5~\% 
(Table~\ref{tab:abu}).
\begin{table}[!h]
\caption{Abundances of the hot gas relative to iron.}
\label{tab:abu}
\centerline{
\begin{tabular}{|lr|lr|}
\hline
Element & Rel. abundance & Element & Rel. abundance \\
\hline
 O & 1.0$\pm$0.3 & S & 1.1$\pm$0.4 \\
 Ne& 0.3$\pm$0.5 & Ar& 2.6$\pm$1.1\\
 Mg& 1.0$\pm$0.3 & Ca& 2.2$\pm$1.4\\
 Si& 0.9$\pm$0.3 & Ni& 0.8$\pm$1.1\\
 Fe& $\equiv$1 & & \\
\hline\noalign{\smallskip}
\end{tabular}
}
\end{table}
The well-determined abundances (O, Mg, Si, S) are all consistent with solar
ratios; the others have in general larger error bars but are not inconsistent
with solar values. Therefore we will adopt solar ratios in this paper, varying
only the global metallicity (iron abundance).

As a next step we fitted spectra in 12 logarithmically spaced annuli with
outer radii ranging between 8--800\arcsec. We also created a set of spectra
corrected for the projection on the sky (assuming spherical symmetry).
The radial temperature and abundance profiles of both sets of spectra
agreed within their error bars, but since the observed (projected) spectra
are less noisy we focus our analysis upon them. However, for the radial
density profile we used the deprojected spectra.

\begin{figure}
\resizebox{\hsize}{!}{\includegraphics[angle=-90]{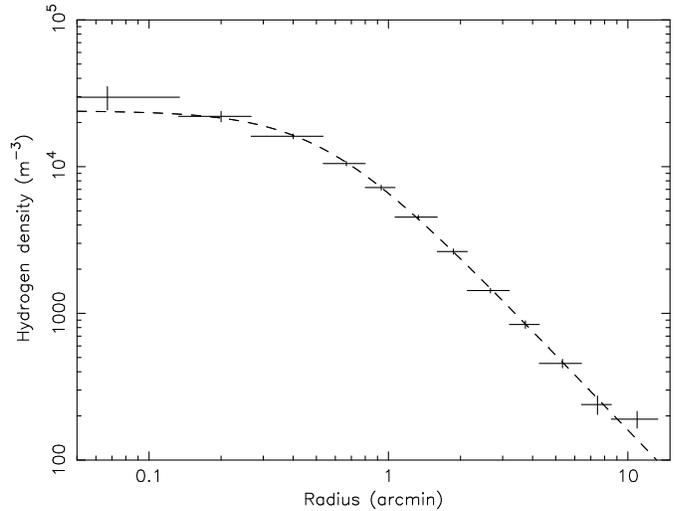}}
\caption{Radial density profile of S\'ersic~159$-$03. The dashed line is the
best-fit $\beta$-model (see text).}
\label{fig:figdens}
\end{figure}

\begin{figure}
\resizebox{\hsize}{!}{\includegraphics[angle=-90]{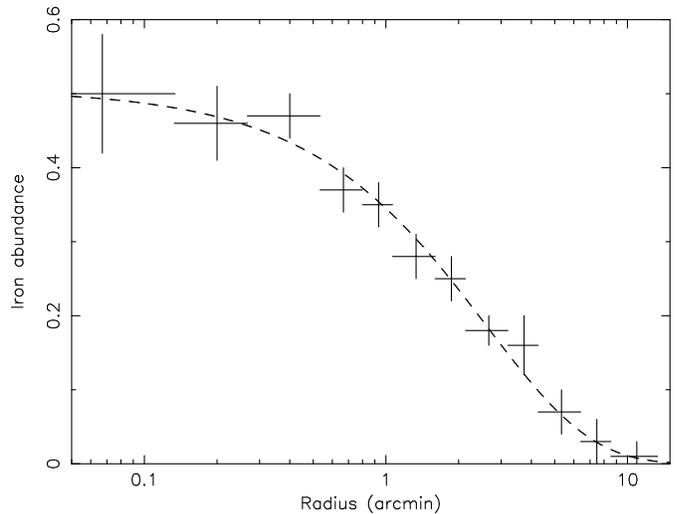}}
\caption{Radial abundance profile of S\'ersic~159$-$03.
The dashed line is the
best-fit exponential model (see text).}
\label{fig:figfe}
\end{figure}

\begin{figure}
\resizebox{\hsize}{!}{\includegraphics[angle=-90]{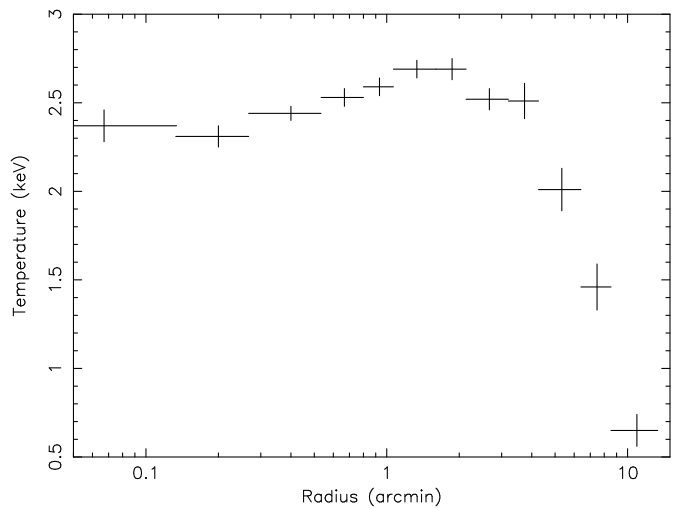}}
\caption{Radial temperature profile of S\'ersic~159$-$03.}
\label{fig:figt}
\end{figure}

The radial density profile is shown in Fig.~\ref{fig:figdens}. The best-fit
$\beta$-model $n_{\rm H} = n_0 [1+(r/a)^2]^{-3\beta/2}$yields a
core radius $a$ of 0.53$\pm$0.06\arcmin\ (47$\pm$5~kpc), a value for
$\beta$ of 0.57$\pm$0.02 and a central hydrogen density $n_0$ of
$(2.40\pm 0.22)\times 10^{4}$~m$^{-3}$. These values are consistent within
the error bars with $\beta$ and $a$ as derived from Rosat HRI observations
(Neumann \& Arnaud \cite{neumann}).
In Fig.~\ref{fig:figfe} we show the radial distribution of iron and
in Fig.~\ref{fig:figt} the temperature distribution. Empirically, the
iron abundance is modelled well by an exponential decay with a 
central abundance of 0.51$\pm$0.04 and a scale height of 2.6$\pm$0.4\arcmin,
although this representation is not unique. Within the central core radius
(0.5\arcmin) the abundance is more or less constant.

S\'ersic~159$-$03\ was reported to have a cooling flow, however Fig.~\ref{fig:figt}
shows only a very modest temperature drop: from $\sim$2.6~keV between
1--6\arcmin\ to about 2.3~keV in the center. The strong iron abundance
gradient and hence the strong Fe-L line complex around 1~keV may have
mimicked partly a cooling flow. Furthermore, a strong temperature
decrease beyond 5\arcmin\ is clearly visible.

We investigated the temperature drop in the center further by fitting
the spectrum within 1\arcmin\ by a hot isothermal model
plus an isobaric cooling flow model (Johnstone et al. \cite{johnstone}).
The temperature of the hot component was frozen to the 1--2\arcmin\ temperature
of 2.69~keV. The best-fit parameters are a mass-deposition rate of
540$\pm$220~M$_{\sun}$/year, metallicity 0.40$\pm$0.02, and a low-temperature
cut-off of 1.37$\pm$0.17~keV. The cooling flow component comprises
about 2/3 of the 2--10~keV flux in this region. The $\chi^2$ of the fit 
was 351 for 308 degrees of freedom (a single temperature fit has only
a slightly worse $\chi^2$ of 373). The low-temperature cut-off of the
cooling flow model is very significant; we could not obtain a satisfactory
fit for very small cut-off temperatures, because these models produce
too much Fe-L emission below 0.9~keV.

\begin{figure}
\resizebox{\hsize}{!}{\includegraphics[angle=-90]{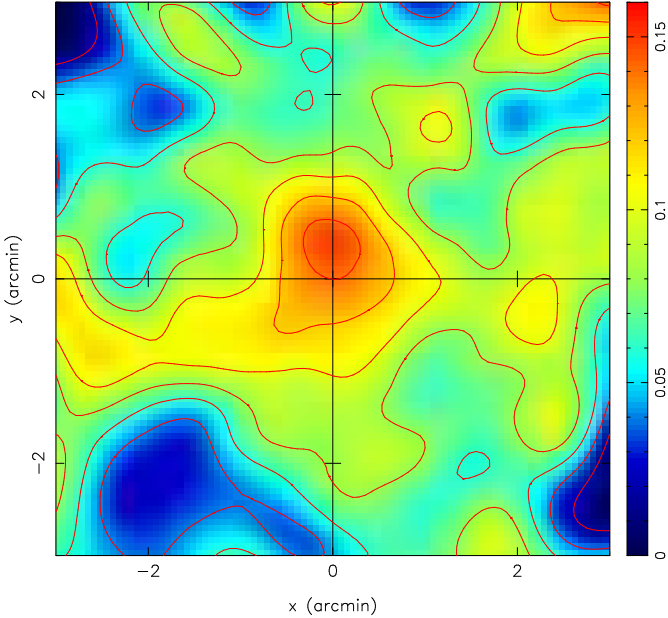}}
\caption{Fe-L equivalent line width in keV. Contours are separated by 0.02~keV.
The statistical uncertainty is about 0.03, 0.05, 0.1 and 0.3 keV at a distance
of 0, 1, 2 and 3\arcmin\ from the core, respectively, for a box of 
16\arcsec$\times$16\arcsec.
}
\label{fig:ew}
\end{figure}

We also studied the azimuthal distribution of the abundances and temperature by
using hardness ratios in selected energy bands.  While the temperature map
(2$-4$~keV / 0.3$-0.5$~keV ratio; not shown) shows no obvious deviations from
the almost isothermal behaviour in the cluster core, the abundance distribution
is not spherical at all.  We produced a map of the equivalent width of the Fe-L
blend between 0.9--1.2~keV as compared to the underlying 0.7--1.3~keV continuum
(Fig.~\ref{fig:ew}).
\begin{figure}
\resizebox{\hsize}{!}{\includegraphics[angle=-90]{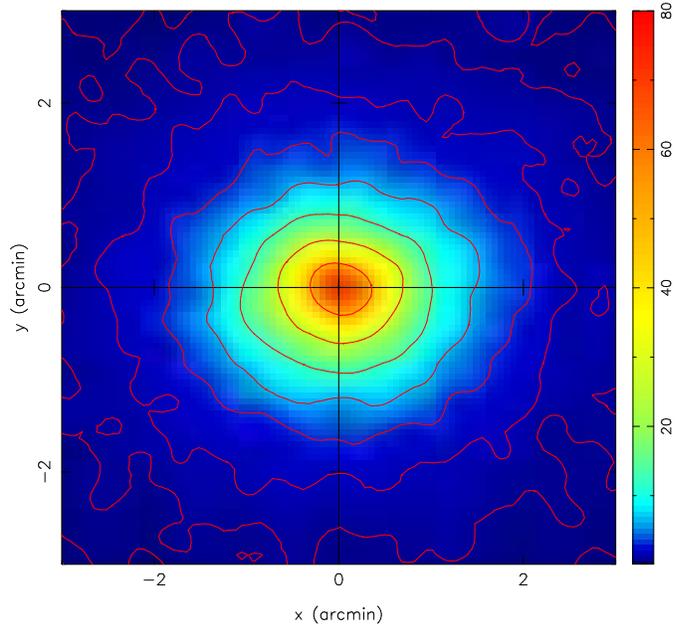}}
\caption{Continuum map for 0.7--0.85 and 1.2--1.3~keV. 
Contours are separated by factors of 2. The field of view is the same
as Fig.~\ref{fig:ew}.
}
\label{fig:cont}
\end{figure}
For the temperature range of interest and accounting for the only minor
variations in temperature based upon the hardness ratio map in the core,
the variations in equivalent width as discussed below translate almost
directly into iron abundance variations.

The map of the equivalent width shows several features. First,
the largest iron abundance is found about 20\arcsec\ to the north
of the brightness peak (compare Fig.~\ref{fig:cont}). Further, there
is an extension out to 2\arcmin\ to the NE of the center, coinciding
more or less with the orientation of the central cD galaxy. Finally,
there is an elongated structure in the EW direction extending several
arcminutes to the east and west, with at least two times more iron than
in other regions at the same distance from the core.

\section{RGS analysis}

\begin{figure}
\resizebox{\hsize}{!}{\includegraphics[angle=-90]{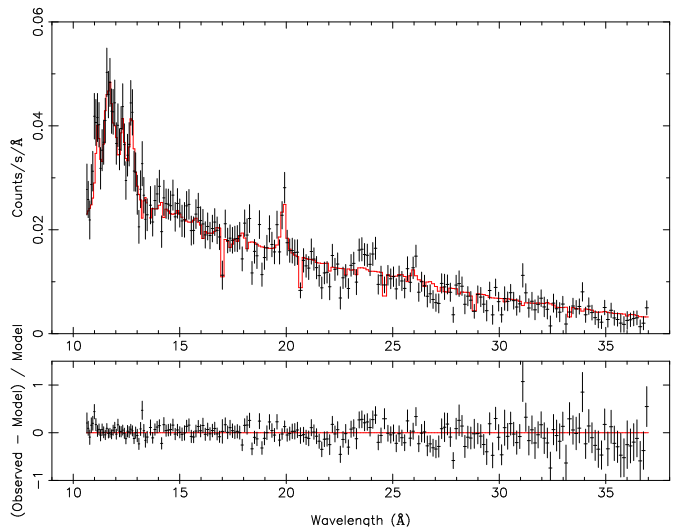}}
\caption{RGS1 spectrum of S\'ersic~159$-$03\ with the best fit 
two component model. Note the \ion{O}{viii} Ly$\alpha$ line at 20.0~\AA;
the lines between 11.2$-$12.8~\AA\ are mainly due to \ion{Fe}{xxiv}, 
\ion{Fe}{xxiii} and \ion{Ne}{x}.}
\label{fig:rgs}
\end{figure}

Firstly, we fitted the spectra with the isothermal collisional ionization
equilibrium model present in the SPEX package (\cite{kmn}).  As in the MOS fits
we took abundances relative to solar (Anders \& Grevesse \cite{anders}).  The
free parameters are the normalization, the temperature and the abundances of
oxygen, neon and iron.  In RGS2, the \ion{O}{viii}~${\rm Ly\alpha}$ line falls
on CCD~4, which failed, and therefore the oxygen abundance is not well
constrained.  The abundances of all the other relevant elements (C, N, Mg, Si,
S, Ar, Ca, Ni) have been coupled to the global metallicity (Fe) consistenly with
the MOS result.  To this model we applied the cosmolgical redshift and galactic
absorption like we did for MOS.  The fit was already satisfactory in both
instruments, but residuals in the region between 13 and 17~\AA\ suggested the
presence of a cooler component.  We added it with the same abundances as the
hotter component and the improvement is $\Delta \chi^2 \simeq 10$.  In
Fig.~\ref{fig:rgs} we show the spectrum obtained with RGS1, in
Table~\ref{tab:aburgs} we reproduce the fitting parameters.  The emission from
the cooler component is about 2\% of the emission from the hotter component,
respectively $5.9\times 10^{35}$~W and $2.5\times 10^{37}$~W.
\begin{table}[!h]
\caption{Fitting parameters and $\chi^2$ for a two components 
collisional equilibrium model. The abundances are relative to solar.
The emission measure is denoted by $Y$.}
\label{tab:aburgs}
\centerline{
\begin{tabular}{|l|c|c|}
\hline
Parameter & RGS1 & RGS2 \\
\hline
$\chi^2 /d.o.f. $ & 220/212 & 229/192 \\
 $Y_1$ (m$^{-3}$)  & $(2.43\pm 0.07)\times 10^{73}$ &  $(3.19\pm 0.10) \times 10^{73}$ \\
 $T_1$ (keV) & 3.0$\pm$0.5 &  2.3$\pm$0.3 \\
 O  & 0.23$\pm$0.07 & 0.16$\pm$0.06 \\
 Ne & 0.8$\pm$0.4 & 0.4$\pm$0.2\\
 Fe & 0.75$\pm$0.19 & 0.45$\pm$0.08\\
 $Y_2$(m$^{-3}$) & $(3.8\pm 1.6)\times 10^{71}$ &  $(3.3\pm 2.5)\times 10^{71}$ \\
 $T_2$ (keV) & 0.77$\pm$0.09 & 0.76$\pm$0.21\\
\hline\noalign{\smallskip}
\end{tabular}
}
\end{table}

The two instruments give slightly different results but consistent within $1
\sigma$, RGS1 gives systematically higher abundances and temperature, this could
be due to residuals problems with the calibration.

The RGS fit is consistent with a model composed by two components:  the dominant
one has $T=2.6 \pm 0.5 $~keV with iron abundance of $0.55 \pm 0.19$ times solar,
a Ne abundance compatible with the Fe one and an oxygen abundance of $0.19 \pm
0.07$ times solar which is small compared with the others.    The
cooler component has a temperature of $0.8 \pm 0.2$~keV and its contribution is
about 2~\% of the total X-ray luminosity.

\section{Discussion}

Our results show that the central part of S\'ersic~159$-$03\ shows very little
temperature structure: the average temperature in the innermost part
is only 10~\% lower than the temperature of the hot gas outside the core.
From imaging data with little spectral information a large mass deposition rate
has been deduced in the past (e.g. 288~M$_{\sun}$/year based upon Einstein IPC
observations (White et al. \cite{white}); 230~M$_{\sun}$/year with the
Rosat PSPC (Allen \& Fabian \cite{allen}). These values will be partly
biased by the strong abundance gradient in the core (Fig.~\ref{fig:figfe}).
Our results with both EPIC and RGS 
are consistent with such a large mass deposition
rate, but only if the emission measure distribution has a low-temperature
cut-off around 1.4~keV. The RGS spectra give upper limits of 1~\% of the
emission measure of the hot gas at any temperature below 1.5~keV.
This upper limit translates into a 0.1--10~keV luminosity of any cool
gas that is less than 20~\% of the predicted luminosity of a 230~M$_{\sun}$/year
cooling flow without a low-temperature cut-off. The absence of cool gas
is similar to what has been found in hotter clusters (A~1835:
Peterson et al. \cite{peterson}; A~1795: Tamura et al. \cite{tamura}).
We refer to those papers for a further discussion of this effect.

Another interesting feature found from our high-resolution RGS spectrum
is the small O/Fe ratio in the core of the cluster: 0.34$\pm$0.17.
This is significantly smaller than the O/Fe ratio of 1.0$\pm$0.3 derived
from EPIC spectral fitting in the outer parts of the cluster.
The oxygen abundance is mainly determined through the strong O~VIII Ly$\alpha$
line. This line is subject to resonance scattering. We estimate the optical
depth of this line to be $\sim$0.5, which yields a relative line
flux depression of no more than 15~\% in the core. We conclude that
there is a real decrease in the O/Fe ratio towards the center of the
cluster. However the absolute oxygen abundance does not differ significantly
between the core (0.19$\pm$0.07, as derived from the RGS) 
and the outer parts (0.19$\pm$0.06,
as derived from EPIC), it is merely the iron abundance
that increases towards the center. Since oxygen is almost totally produced
by type II supernovae and iron mostly by type Ia supernovae, we conclude
that the core of the cluster has been relatively rich in type Ia supernovae,
while the type II supernovae are more or less uniformly distributed over the
cluster. The oxygen from type II supernovae may have been ejected
in protogalactic winds (Larson \& Dinerstein \cite{larson}). 
The dominance of type Ia supernovae in the center may be due to
the dominance of elliptical galaxies with an old stellar population
in the center of the cluster. Due to ram pressure stripping these galaxies
can loose their metals to the intracluster medium (Gunn \& Gott \cite{gunn}).

The structure seen in the Fe-L equivalent width map (Fig.~\ref{fig:ew})
in the central cD galaxy (within 30\arcsec\ from the core) indicates
that the iron is distributed inhomogeneously in the core. A possible
explanation might be a recent merger of a gas-rich galaxy in the NE
part of the cD galaxy. Iron is not expected to diffuse over large distances
during the life-time of a cluster. Therefore the elongated iron-rich structure
seen in the EW direction extending out to at least 3\arcmin\ (0.3~Mpc)
towards the east probably reflects the initial star formation in the
early evolution phase of the cluster. The additional amount of iron in
this structure is consistent with the total amount of interstellar
iron from a solar-composition galaxy with $\sim 10^{10}$~M$_{\sun}$\ of gas.
Rosat PSPC images show indications for
an extended, elongated structure out to at least half a degree to the east of
the core, with a width of a few arcminutes. The characteristic density
in this structure is of the order of 100--200~m$^{-3}$, similar to the
average cluster density in our outermost annulus (Fig.~\ref{fig:figdens}).
Perhaps these structures are associated with a supercluster; unfortunately
the region around S\'ersic~159$-$03\ is poorly studied, so we cannot confirm this.

A possible cause of the strong temperature drop near the outer part of the
cluster might also be associated with the transition from cluster to
supercluster.  Temperature drops by a factor of 2 or more and an associated
metallicity gradient, on slightly larger spatial scales than in
S\'ersic~159$-$03\ have been found e.g.  in the A 3562/Shapley supercluster
(Kull \& B\"ohringer \cite{kull}; Ettori et al.  \cite{ettori}).  Numerical
models of cluster mergers (Ricker \cite{ricker}) show that after the merger
there can be large-scale temperature gradients in the outer parts of a cluster.
This is due to a post-merger accretion shock caused by gas falling back after
the merging of the cores.  The temperature can drop by an order of magnitude at
10 core radii (about 5\arcmin\ in S\'ersic~159$-$03), qualitatively similar to
what we find.

\begin{acknowledgements}
This work is based on observations obtained with XMM-Newton, an ESA science 
mission with instruments and contributions directly funded by 
ESA Member States and the USA (NASA).
The Laboratory for Space Research Utrecht is supported
financially by NWO, the Netherlands Organization for Scientific
Research. 

\end{acknowledgements}

\end{document}